\documentstyle[ 12pt, psfig]{article}

\def\refs{\leftskip=.3truein\parindent=-.3truein}
\def\unrefs{\leftskip=0.0truein\parindent=20pt}

\title{A Brief History of Astronomical Brightness Determination Methods
at Optical Wavelengths}

\author {Kevin Krisciunas}
\date { }

\begin{document}
\maketitle


\begin {abstract}
In this brief article I review the history of astronomical photometry, touching on
observations made by the ancient Chinese, Hipparchus and Ptolemy,
the development of the concept (and definition) of magnitude,
the endeavors of Argelander and Z\"{o}llner, work at Harvard at
the end of the 19th century, and the development of photography,
photomultipliers, and CCD's and their application to astronomy.
\end {abstract}

\vspace {3 mm}

\parindent = 15 mm
{\em ...this brave o'erhanging firmament, this majestical 
roof fretted with golden fire....}

\begin {center}
{\em Hamlet}, II, ii, 308-310
\end {center}

\vspace {3 mm}

\parindent = 9 mm

The following introduction to photometry served as Chapter 1 to my
recent University of Washington Dissertation; see Krisciunas (2001)
for a summary.  Chapter 2 (Krisciunas, Margon, \& Szkody 1998) dealt
with the identification of objects of interest, in particular RR
Lyrae stars, carbon stars, asteroids, and cataclysmic variables,
using the photometric system being used for the Sloan Digital Sky
Survey.  Chapter 3 involved the confirmation of RR Lyrae
candidates from SDSS as {\em bona fide} RR Lyrae stars; light curves
of six of these can be found in Ivezic et al. (2000).  Chapters 4 and
5 (Krisciunas et al. 2000, 2001)  dealt with Type Ia supernovae and
the use of optical and infrared photometry to determine the
extinction of these objects attributable to dust along the line of
sight.

\vspace{1 cm}

The first systematic observations of the heavens which can be
considered ``scientific'' in any sense were carried out by
the Chinese as early as the 14th century BC (Needham \& Ling 1959, p. 459).
During the Tang Dynasty (618$-$904 AD) the Chinese Astronomical Bureau
consisted of 50 ranked officials directing as many as 500 to 900 personnel
(Deane 1989, p. 139).  For many centuries it was typical that their
Astronomical Bureau employed three to four dozen ``astronomers''.  The
prime motivation was revision of their luni-solar calendar and the
designation of auspicious days for the carrying out of many state
rituals.  Any observations of the night sky necessarily involved knowledge
of the stellar asterisms (i.e. constellations).  

By the time Galileo was observing the heavens with a small refractor at
the beginning of the 17th century, the Chinese had been recording
celestial phenomena for nearly 3000 years.  Many of those observations are
of use to modern astronomers.

Chinese historical records contain observations of sunspots as far back as
the first century BC (Needham \& Ling 1959, p. 434), possible naked eye
detections of Jupiter's moon Ganymede in 364 BC (Xi 1981), plus
observations of comets, novae, supernovae, and other variable stars
(Needham \& Ling 1959, p. 423; Clark \& Stephenson 1977).  The
``historical supernovae'' occurred in AD 185, 386, 393, 1006, 1054 (what
is now the Crab Nebula in Taurus), 1181, 1572 (Tycho's Supernova), and
1604 (Kepler's Supernova).\footnote[1]{As a final ``historical
supernova'' we might consider Flamsteed's star number 3 Cas, 
observed by him in 1680.  It is very likely the
progenitor of the Cas A supernova remnant (Ashworth 1980).}

Clark \& Stephenson (1977) discuss at length the Chinese records and try
to identify the locations of the progenitors of the supposed supernovae
using coordinates of known supernova remnants.  Schaefer (1993, 1995,
1996) attempted to determine the peak brightnesses of the possible Type Ia
supernovae (those of 185, 1006, 1572, and 1604) and found problems with
the interpretation of the historical records, concluding that SN 185 might
not even have been a supernova.  For example, SN 185 was described to be
``as large as half a mat''.  Was this a description of its brightness or
an admission that it was a non-stellar object such as a comet?

The concept of {\em stellar magnitudes} is at least as old as Ptolemy's
{\em Almagest} (ca. 137 AD).  Ptolemy gives a list of 1022 stars arranged
in 48 constellations, with ecliptic coordinates and a magnitude for each
star. Ptolemy's catalogue was based, or largely borrowed (with an
incorrect value for precession) from the star catalogue of Hipparchus (ca.
137 BC).  The literature on Ptolemy's star catalogue is quite extensive
(see Evans 1998, pp. 264-274), and we will not cover the subject here.  
Suffice it to say that the our sensory organs and the brain perceive
stimuli (such as light, sound, and taste)  proportional to the logarithm
of the stimulus.  This is known as the Weber-Fechner psychophysical law
(Herrmann 1984, p. 73; Hearnshaw 1996, p. 73).  Stars ``of the first
order'' were called ``first magnitude'' by Ptolemy (or Hipparchus before
him).  The faintest stars visible to the unaided eye were designated to be
of sixth magnitude.\footnote[2]{Weaver (1946, p. 224) suggests that the
choice of {\em six} magnitudes was probably connected with the use of the
sexagesimal system for measuring time and angles.}

The magnitude scale was defined in 1856 by Norman Pogson (Hearnshaw 1992)
as follows: one star that has an apparent brightness 100 times that of a
second star is by definition five magnitudes brighter.  Thus, each
magnitude corresponds to a ratio of luminosities equal to the fifth root
of 100 (roughly 2.512).  Even with this definition, there is always the
problem of a zero point.  One can read about stellar magnitudes, but until
one takes a star atlas outside at night and sees what constitutes the
brightness of an actual third, or second magnitude star, it is just an
abstract number.

While observers through the late nineteenth century might agree on the
magnitudes of the stars visible to the unaided eye, large systematic
differences could be obtained with respect to stars seen in the telescope
eyepiece.  For example, Pogson found that Wilhelm Struve's magnitude 11.9
corresponded with 17.9 of John Herschel, but both correspond to magnitude
15 on today's scale (Hearnshaw 1992).  Still, magnitudes are excellent
{\em relative} units, since most observers can agree as to which of two
stars is brighter.

The ability of the eye to discern differences of brightness allows
one to discover and monitor the light curves of variable stars.  
This is usually called the ``Argelander step method'' after the
German astronomer F. W. A. Argelander (1799$-$1875).  Members of
organizations such as the American Association of Variable Star
Observers (AAVSO) use charts prepared from photographic or
photoelectric photometry to make visual light curves of many
thousands of variable stars, a task of great benefit to professionals
needing to know, for example, when a particular dwarf nova has
reached outburst so that they might aim an orbiting ultraviolet
satellite at it (see Fig. 1).

One of the most impressive projects of the 19th century was the {\em
Bonner Durchmusterung}, or {\em BD}, which contained the positions and
magnitudes (to the nearest 0.1 mag) of 324,198 stars north of declination
$-2^{\rm o}$ (Ashbrook 1980).  It was carried out with a 3.1-inch
refractor at the Bonn Observatory under Argelander's direction and was
based on visual observations of stars transiting a reticle in the
eyepiece.  The telescope was set at a
specific declination and the stars were allowed to drift through
the field. The observer called off the magnitude and relative
declination of each star transiting, and a recorder wrote this down and
noted the sidereal time.

The {\em BD} was published between 1859 to 1862 and was then extended to
southern skies.  133,659 stars with declinations from $-2^{\rm o}$ to
$-22^{\rm o}$ were added by 1886, and the {\em Cordoba Durchmusterung},
carried out in Argentina and published between 1892 and 1914, added
another 578,802 stars with declinations between $-22^{\rm o}$ and 
$-62^{\rm o}$.  It was extended to the South Celestial Pole by 1932.

After its invention by L. J. M. Daguerre (1787$-$1851) and J. N. Niepce
(1765$-$ 1833), photography developed slowly (Herrmann 1984, p. 81; 
Krisciunas 1988, p. 127).  The first photograph of a
star was obtained in 1850 by J. A. Whipple, working under the direction of
W. C. Bond (1789$-$1859).  They used the Harvard College Observatory
15-inch refractor (Jones \& Boyd 1971, p. 76).  However, stars fainter
than first magnitude could not be registered, so Bond suspended further
experiments until 1857, by which time photographic plates were sensitive
enough to register sixth or seventh magnitude stars with the 15-inch.  

A great leap in sensitivity was achieved in 1874 with the invention of dry
gelatino-bromide plates by W. Abney (1843$-$1920).  By 1880 the American
H. Draper (1837$-$1882) was able to photograph the Orion Nebula (Holden
1882, p. 226).  Large international projects such as the {\em Carte du
Ciel}, begun in 1887, and the beautiful photographs of nebulae by Keeler
at Lick Observatory, using the 36-inch Crossley reflector, showed that
astronomical photography had come of age (Krisciunas 1988, pp. 73, 151).  
Details of photographic photometry are discussed by Weaver (1946).

In the 1850's J. K. F. Z\"{o}llner (1834$-$1882) developed the first
practical astronomical photometer, which embodied an application of
the polarization of light.  In the Z\"{o}llner photometer the focal
image of a real star was compared with the focal image of an
artificial star made visible in the same field of view.  The
brightness of the artificial star could be adjusted by the relative
position angles of two Nicol prisms until it matched the real star,
and the orientation of the prisms allowed the magnitudes of the stars
to be derived (Weaver 1946, p. 214; Herrmann 1984, p. 73; Hearnshaw
1996, p. 61).  In 1861 Z\"{o}llner published data on 226 stars and
soon after supplied 22 such photometers to observatories in Russia,
the United Stares, England, Holland and other countries.

After E. C. Pickering (1846$-$1919) became Director of Harvard College
Observatory in 1877 he experimented with various photometers built on the
polarization principle, but then opted for a new design. Pickering's
meridian photometer consisted of a horizontal telescope with two
objectives of the same diameter and approximately same focal length.  One
objective was used to measure a standard star near the North Celestial
Pole, while the other was used to observe some other target star.  As in
the Z\"{o}llner photometer, this system relied on the observer's ability
to judge when two light sources were of equal apparent brightness.  Of
course, this could lead to systematic differences from observer to
observer, especially if the stars were very red, but measurements could be
made to $\pm$ 0.1 mag.  It was on the basis of observations with a
meridian photometer that Pickering and his collaborators produced the {\em
Harvard Revised Photometry},\footnote[3]{This forms the basis of the
various editions of the {\em Bright Star Catalogue}.  The ``{\em HR}''
numbers still in use today refer to Pickering's numbering scheme.}
comprising 9110 stars brighter than magnitude 6.5 and ranging from the
North Celestial Pole to the South Celestial Pole.

J. Stebbins (1878$-$1966) began experimenting with a selenium photocell at
the University of Illionis and published the first photometric light curve
of the moon's brightness as a function of phase (Stebbins \& Brown 1907).
Three years later (Stebbins 1910) he published the most accurate set of
photometric observations in existence at that time, a light curve of the
eclipsing binary Algol ($\beta$ Persei).  This was the first demonstration
of the secondary minimum.  His data had probable errors as low as $\pm$
0.006 mag.  With the adoption of the Pogson scale and the development
of photographic and photoelectric photometry, by the beginning of the
20th century astronomers had a true system of brightness measurements $-$
a far cry from Ptolemy's approximate magnitude scale.

Stebbins and two physicists at the University of Illinois, J. Kunz and W.
F. Schulz, developed photometers with photoelectric cells as early as
1913.  P. Guthnick in Berlin, and H. O. Rosenberg and F. Meyer in
T\"{u}bingen were also pioneers in the field of photoelectric photometry
at this time (Weaver 1946, p. 507 ff.; Hearnshaw 1996, p. 193 ff.). With
the invention of the {\em photomultiplier tube} after 1945, however,
photoelectric photometry came of age and replaced photographic photometry
as the principal method of making brightness measurements of stars
(Hearnshaw 1996, p. 411 ff.).  An example of a photoelectric light curve 
is given in Fig. 2.

A photoelectric photometer allows one to measure the light of a small
patch of sky by means of a diaphragm, a Fabry (field) lens, and a set
of colored glass filters.  One reimages the light of the telescope's
primary mirror on the photocathode of the photomultiplier tube, whose
output is amplified.  One can either operate the system by recording
a DC signal on a strip chart recorder, or one can build pulse
counting electronics, which allows one to detect {\em individual}
photons. One of the advantages such a device is that one generally
uses a diaphragm much larger than the seeing disk at one's site (20
to 30 arcsec might be typical); changes in the astronomical seeing do
not affect the data in any significant way.  However, because one is
measuring one star-plus-sky at a time, one must take separate sky
readings, and the number of sky position/filter integrations one can
take on a clear night is usually limited to a few hundred.  Since
many observations are of standard stars, one might only obtain a few
dozen program observations on any night.

The most widely used photometric system is the UBV system of Johnson \&
Morgan (1953), which is based on three carefully selected broad-band
filters with effective wavelengths of 360, 440, and 550 nm and the
photoelectric response of a CsSb (S-4) photosurface such as that of the
RCA 1P21 photomultiplier tube.  R-band (700 nm) and I-band (900 nm)
filters have been used as well (Johnson 1966), but a more red-sensitive
photosurface such as the S-20 needs to be used to improve sensitivity at
those wavelengths (Walker 1987, p. 217).

Many other photometric systems have been developed; see Hearnshaw
(1996, p. 434 ff.) for a summary and further references.  One of the
most widely used is the system devised by B. Str\"{o}mgren
(1908$-$1987), which allows the determination of stellar radii,
surface gravities, effective temperatures, and absolute magnitudes
(Balona \& Shobbrook 1984, Moon 1984, Moon \& Dworetsky 1985).
Another system which will become widely used in the near future
is being used for the Sloan Digital Sky Survey (Gunn \& Knapp 1993; 
Fukugita et al. 1996).

One of the advantages of photomultipliers is their greater sensitivity
compared to photographic methods. The {\em quantum efficiency} $-$ or QE,
the efficiency of a light sensitive element in detecting photons $-$ of a
photographic plate is typically one percent, though hypersensitized
emulsions can achieve QE's of 4$-$5 percent (Walker 1987, p. 266; Rieke
1994, p. 39).  A photomultiplier tube can achieve a maximum QE in excess
of 10 percent (Walker 1987, p. 217), and cooling a photomultiplier tube
with dry ice or thermoelectrically will decrease the ``dark current''
substantially, resulting in increased sensitivity because of lower noise.

The modern imaging device of choice is the charge-coupled device, or CCD
(McLean 1989; Janesick \& Elliott 1992), which was first introduced to the
world in 1970.  When light illuminates a CCD, four processes occur: 1)
charge generation; 2) charge collection; 3) charge transfer; and 4) charge
detection.  The {\em charge generation} occurs as a result of the
the production of electron-hole pairs in the light sensitive chip.
The {\em charge collection} takes place in
the nearest discrete collecting sites, or pixels.  The collection sites
are defined by an array of electrodes, called gates.  The {\em charge
transfer} results from manipulating the voltages on the gates so that the
electrons move down vertical registers from one pixel to the next, like a
bucket brigade.  A horizontal register at the end of each column collects
the charges in a serial manner and transports them to an on-chip
amplifier.  Finally, the {\em charge detection} is accomplished when the
individual charge packets are converted to output voltages.

There are two significant advantages of CCD's over photomultiplier tubes.
First of all, CCD's can achieve QE's in excess of 70 percent, allowing
much fainter objects to be detected without the construction of a larger
telescope.  Secondly, because a CCD is an {\em array} detector, one can
observe multiple objects in the same frame {\em and} measure the sky
brightness as well.  This alone increases observing efficiency by a factor
of three or more.  Observations of crowded fields, such as in globular
clusters, are nearly impossible with photomultiplier tubes, but are now
routine with CCD's, providing one is using an appropriately sophisticated
data reduction package such as DoPhot (Schechter, Mateo, \& Saha 1993)
or {\sc daophot} (Stetson 1987, 1990).

One difference between photomultiplier tubes and CCD's is that the former
typically have peak QE in the B-band, while the latter have peak QE
in the R-band.  The development of special anti-reflection coatings
and the use of back-illuminated CCD's has not only boosted their QE's,
but has made them more equally sensitive over their whole range of
wavelengths.

\newpage

\begin {center}
{\bf References}
\end {center}

\refs

Ashbrook, J. 1980, {\em Sky and Telescope}, {\bf 59}, 300

Ashworth, W. B. 1980, {\em J. for the History of Astronomy}, {\bf 11}, 1

Balona, L., \& Shobbrook, R. R. 1984, {\em Monthly Notices Royal Astron.
Soc.}, {\bf 211}, 375

Bester, M., Danchi, W. C., Hale, D., Townes, C. H., Degiacomi, C. G.,
M$\acute {\rm e}$karnia, D., \& Geballe, T. R. 1996, {\em Astrophys. J.},
{\bf 463}, 336

Clark, D. H., \& Stephenson, F. R. 1977, {\em The Historical Supernovae}
(New York: Pergamon Press)

Deane, T. E. 1989, {\em The Chinese Imperial Astronomical Bureau: Form and
Function of the Ming Dynasty {\rm Qintianjian} from 1365 to 1627}, Univ. of
Washington Dissertation

Dupree, A. K., Baliunas, S. L., Guinan, E. F., Hartmann, L.,
Nassiopoulos, G. E., \& Sonneborn, G. 1987, {\em Astrophys. J.},
{\bf 317}, L85

Evans, J. 1998, {\em The History and Practice of Ancient Astronomy}
(New York and Oxford: Oxford Univ. Press)

Fukugita, M., Ichikawa, T., Gunn, J. E.,
Doi, M., Shimasaku, K., \& Schneider, D. P. 1996, {\em Astron. J.},
{\bf 111}, 1748

Gunn, J. E., \& Knapp, G. R. 1993, in {\em Sky Surveys:
Protostars to Protogalaxies}, B. T. Soifer, ed. (San Francisco:
Astronomical Society of the Pacific), ASP Conf. Series, {\bf 43}, p.
267

Hearnshaw, J. B. 1992, {\em Sky and Telescope}, {\bf 84}, 492

Hearnshaw, J. B. 1996, {\em The Measurement of Starlight: Two centuries
of astronomical photometry} (Cambridge: Cambridge Univ. Press)

Herrmann, D. B. 1984, {\em The History of Astronomy from Herschel to
Hertzsprung}, translated by Kevin Krisciunas (Cambridge: Cambridge Univ.
Press)

Holden, E. S. 1882, {\em Monograph of the Central Parts of the
Nebula of Orion}, Washington: Government Printing Office, Washington
Astronomical Observations for 1878, Appendix I

Ivezic, Z., et al. 2000, {\em Astron. J.}, {\bf 120}, 963 (astro-ph/0004130)

Janesick, J., \& Elliott, T. 1992, in {\em Astronomical CCD
Observing and Reduction Techniques}, S. B. Howell, ed. (San Francisco:
Astronomical Society of the Pacific), ASP Conf. Series, {\bf 23}, p. 1

Johnson, H. L., \& Morgan, W. W. 1953, {\em Astrophys. J.}, {\bf 117}, 313

Johnson, H. L. 1966, {\em Annual Review of Astronomy and Astrophysics}, 
{\bf 4}, 193

Jones, B. Z. \& Boyd, L. G. 1971, {\em The Harvard College Observatory:
The first four directorships, 1839$-$1919} (Cambridge,
Massachusetts: Harvard Univ. Press)

Krisciunas, K. 1988, {\em Astronomical Centers of the World} (Cambridge:
Cambridge Univ. Press)

Krisciunas, K., Margon, B., \& Szkody, P. 1998, {\em Publ. Astron.
Soc. Pacific}, {\bf 110}, 1342 (astro-ph/9808093)

Krisciunas, K., Hastings, N. C., Loomis, K., McMillan, R., Rest, A.,
Riess, A. G., \& Stubbs, C. 2000, {\em Astrophys. J.}, {\bf 539}, 658
(astro-ph/9912219)

Krisciunas, K. 2001, {\em Publ. Astron. Soc. Pacific}, {\bf 113}, 121

Krisciunas, K., Phillips, M. M., Stubbs, C., Rest, A., Miknaitis, G.,
Riess, A. G., Suntzeff, N. B., Roth, M., Persson, S. E., \& Freedman,
W. L. 2001, {\em Astron. J.}, in press (astro-ph/0106088)

McLean, I. S. 1989, {\em Electronic and Computer-Aided Astronomy: from
eyes to electronic sensors} (New York: Halsted Press)

Moon, T. T. 1984, {\em Monthly Notices Royal Astron. Soc.}, {\bf 211}, 21{\sc P}

Moon, T. T., \& Dworetsky, M. M. 1985, {\em Monthly Notices Royal Astron.
Soc.}, {\bf 217}, 305

Needham, J., \& Ling, W. 1959, {\em Science and Civilisation in China},
vol. 3, {\em Mathematics and the Sciences of the Heavens and the Earth}
(Cambridge: Cambridge Univ. Press)

Rieke, G. H. 1994, {\em Detection of Light: from the ultraviolet
to the submillimeter} (Cambridge: Cambridge Univ. Press)

Schaefer, B. E. 1993, {\em Publ. Astron. Soc. Pacific}, {\bf 105}, 1238

Schaefer, B. E. 1995, {\em Astron. J.}, {\bf 110}, 1793

Schaefer, B. E. 1996, {\em Astrophys. J.}, {\bf 459}, 438

Schechter, P., Mateo, M., \& Saha, A. 1993,
{\em Publ. Astron. Soc. Pacific}, {\bf 105}, 1342

Stebbins, J., \& Brown, F. C. 1907, {\em Astrophys. J.}, {\bf 26}, 326

Stebbins, J. 1910, {\em Astrophys. J.}, {\bf 32}, 185

Stetson, P. 1987, {\em Publ. Astron. Soc. Pacific}, {\bf 99}, 191

Stetson, P. 1990, {\em Publ. Astron. Soc. Pacific}, {\bf 102}, 932

Walker, G. 1987, {\em Astronomical Observations} (Cambridge: Cambridge
Univ. Press)

Weaver, H. 1946, {\em Popular Astronomy}, {\bf 54}, 211, 287, 339, 389,
451, 504

Xi, Z.-Z. 1981, {\em Chinese Astronomy and Astrophysics}, {\bf 5}, 242
(or see {\em Sky and Telescope}, February 1982, {\bf 63}, 145)

\unrefs

\begin{figure*}
\psfig{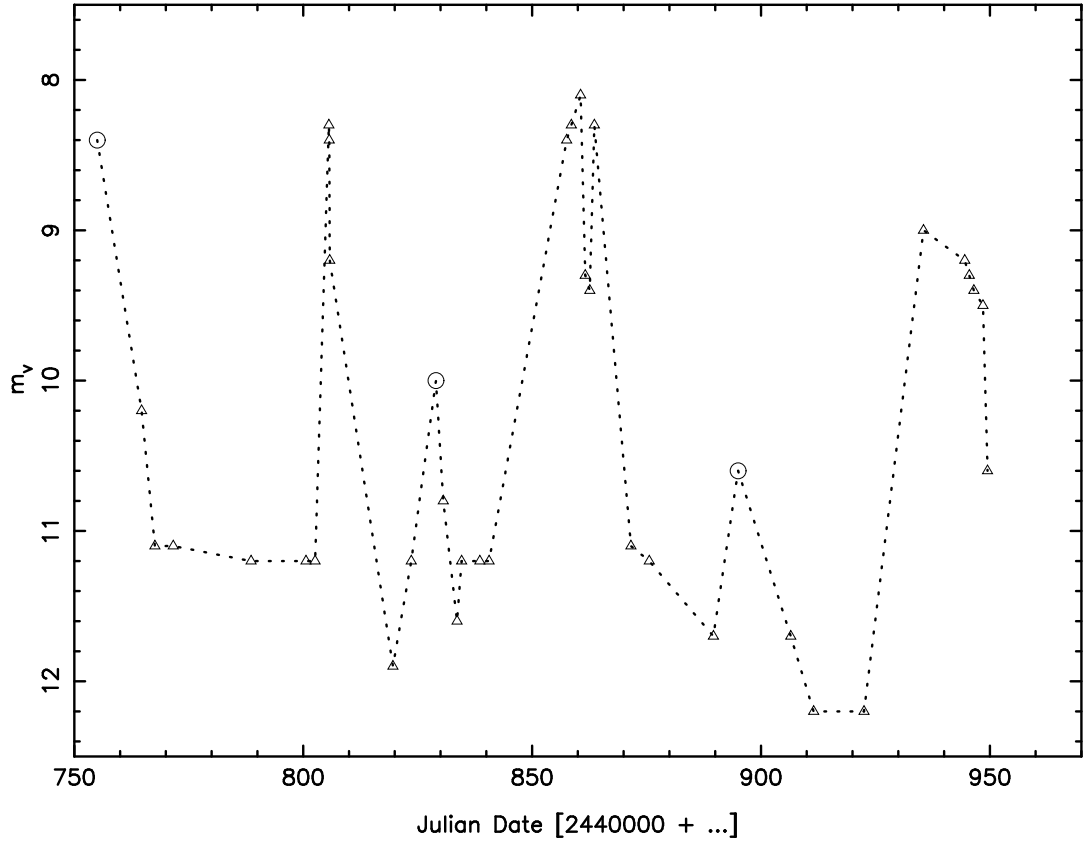}
\caption[Visual light curve of SS Cygni]
{Visual light curve of the dwarf nova
SS Cygni from June to December 1970, using
charts provided by the AAVSO. Triangles are data by K. Krisciunas, while
large open circles are the approximate mean values of other data
reported to the AAVSO.  A visual estimate is typically accurate
to $\pm$ 0.1 to 0.2 mag.}
\end{figure*}

\begin{figure*}
\psfig{figure=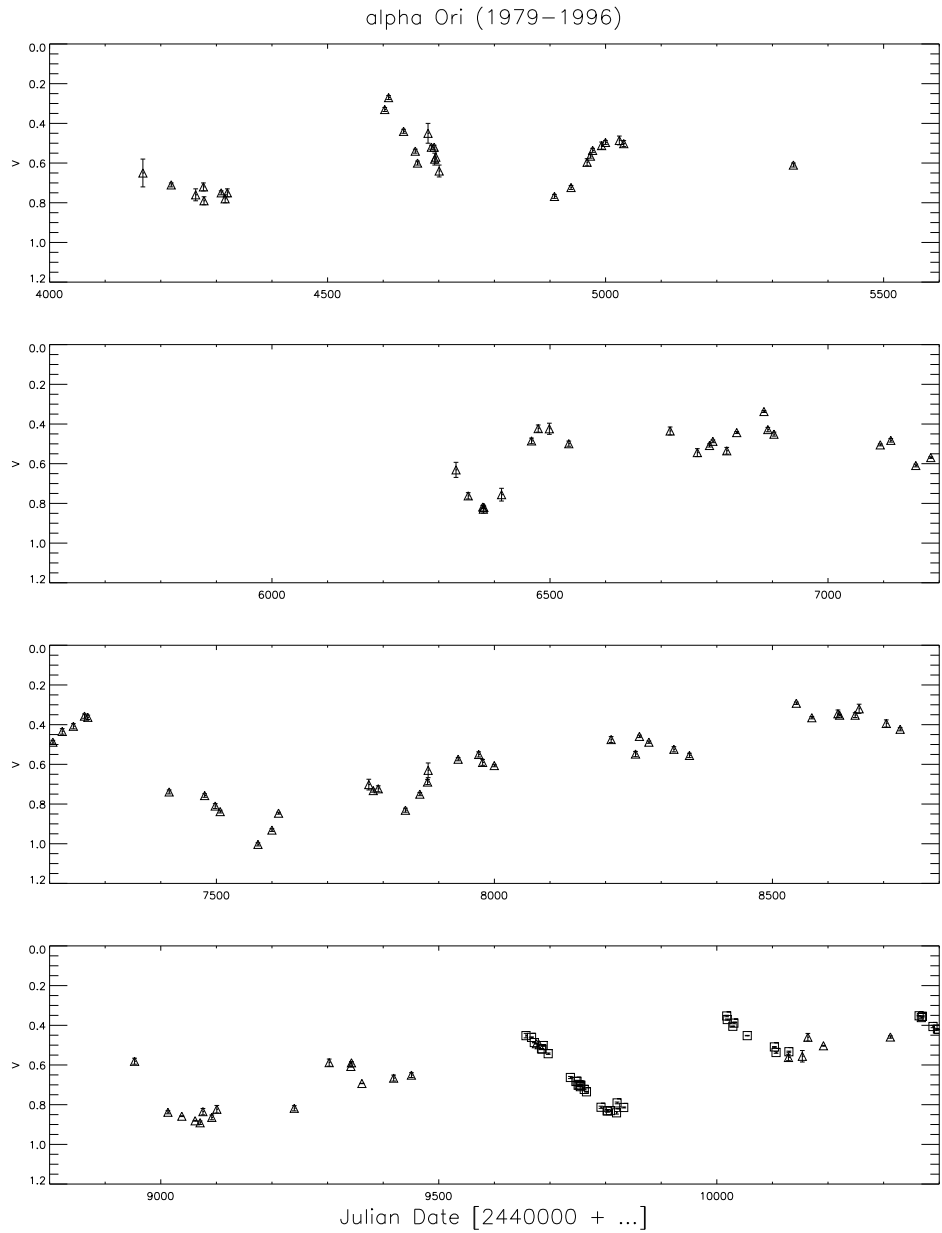,height=16cm,angle=0}
\caption[Photoelectric light curve of $\alpha$ Orionis] {Photoelectric
V-band light curve of $\alpha$ Orionis (Betelgeuse) from
21 October 1979 to 11 November 1996 UT, with data by K. Krisciunas (triangles)
and K. Luedeke (small squares). 
These data have been used by Dupree et al. (1987) to correlate with
chromospheric modulation of the star, and by Bester et al. (1996) to
correlate with interferometric measurements of the diameter of Betelgeuse
at 11 $\mu$m.}
\end{figure*}

\end{document}